\documentclass[fleqn,10pt]{wlscirep}
\usepackage[utf8]{inputenc}
\usepackage[T1]{fontenc}
\usepackage[normalem]{ulem}
\usepackage{color}
\usepackage{graphicx}
\usepackage{float}
\newcommand{\blue}[1]{\textcolor{blue}{#1}}

\newcommand{\stk}[1]{\ifmmode\text{\sout{\ensuremath{#1}}}\else\sout{#1}\fi}

\title{Learning and criticality in a self-organizing model of connectome growth}

\author[1,*]{Michelle T. Cirunay}
\author[2,3]{Rene C. Batac}
\author[1]{G\'eza \'Odor}

\affil[1]{Institute of Technical Physics and Materials Science, HUN-REN Centre for Energy Research, P.O. Box 49, H-1525, Budapest, H-1525, Hungary}
\affil[2]{Department of Physics, College of Science, De La Salle University, 2401 Taft Ave., Manila, 0922, Philippines}
\affil[3]{Dr. Andrew L. Tan Data Science Institute, De La Salle University, 2401 Taft Ave., Manila, 0922, Philippines}

\affil[*]{michelle.cirunay@ek.hun-ren.hu}

 \begin{abstract}
	\noindent\textbf{The exploration of brain networks has reached an important milestone as relatively large and reliable information has been gathered for connectomes of different species. Analyses of connectome data sets reveal that the structural length\cite{RavaszNEU2013} and the distributions of in- and out-node strengths\cite{GastnerOdorSCIREP2016,CirunayEtAlPRR2025} follow heavy-tailed lognormal statistics, while the functional network properties exhibit powerlaw tails\cite{ChialvoBakNEUROSCI1999}, suggesting that the brain operates close to a critical point where computational capabilities\cite{LegensteinMaassNEURALNETW2007} and sensitivity to stimulus is optimal\cite{KinouchiCopelliNATUREPHYS2006}. Because these universal network features emerge from bottom-up (self-)organization, one can pose the question of whether they can be modeled via a common framework, particularly through the lens of criticality of statistical physical systems. Here, we simultaneously reproduce the powerlaw statistics of connectome edge weights and the lognormal distributions of node strengths from an avalanche-type model with learning that operates on baseline networks that mimic the neuronal circuitry. We observe that the avalanches created by a sandpile-like model\cite{BakEtAlPRL1987} on simulated neurons connected by a hierarchical modular network (HMN)\cite{KaiserFRONTIERS2010,OdorEtAlSCIREP2015} produce robust powerlaw avalanche size distributions with critical exponents of 3/2 characteristic of neuronal systems\cite{BeggsJNEURO2003}. Introducing Hebbian learning, wherein neurons that `fire together, wire together,' recovers the powerlaw distribution of edge weights and the lognormal distributions of node degrees, comparable to those obtained from connectome data\cite{CirunayEtAlPRR2025}. Our results strengthen the notion of a critical brain, one whose local interactions drive connectivity and learning without a need for external intervention and precise tuning\cite{Beggs2022}.  }
 \end{abstract}
\begin{document}

\maketitle



The association between form and function in the complex neuronal circuitry continues to be an open and exciting problem, one that is now being pursued with the aid of empirical data on actual connectome networks. Recently, global weight distributions of the human white matter\cite{OdorPRE2016} and neuronal-level fruit fly brain\cite{OdorEtAlPRR2022} have been found to exhibit powerlaw (PL) tails. A more detailed network analysis, involving more full or partial brain networks\cite{CirunayEtAlPRR2025} has strengthened this PL tail observation for global weights, in contrast with node strength (degree) distributions, which can be best fitted by heavy-tailed, typically lognormal (LN) functions. The existence of robust PL statistics raises the possibility of analyzing these systems from the lens of statistical physics, particularly through a criticality framework.

The global weight distribution can be the consequence of the neuron dynamics via the Hebbian learning of long-term synaptic (LTS) mechanism. In a previous work\cite{CirunayEtAlPRR2025}, a possible scaling argument has been proposed to describe the observed connectome weight PL distribution via the ``firing together and wiring together'' mechanism. In particular, the universally observed exponent 3 decay can be derived by a relation to the mean-field criticality of simple models of statistical physics. Here, as a plausible framework for modeling the observed universal statistics, we provide a critical system, a self-organized one, which has extensively been used as a toy model to support brain criticality. The brain criticality hypothesis\cite{ChialvoBakNEUROSCI1999,ChialvoNATUREPHYS2010} has been observed in experiments\cite{BeggsJNEURO2003,PalmaEtAlPNAS2013} and supported by many theoretical models\cite{MunozRMP2018}. From a discrete modeling perspective, critical behavior is often associated with the avalanche-like dynamics characteristic of self-organized criticality (SOC)\cite{BakEtAlPRL1987,HaiEtAlPRL2013,HesseGrossFRONTSYSTNEUROSCI2014}, the paradigm model of which is the sandpile. In a sandpile model, local relaxations lead to cascades of nearest-neighbor redistributions. Previous models replicated empirical time signatures of neuronal activity using nearest-neighbor avalanches similar to the sandpile operating in lattice and small-world networks\cite{DeArcangelisEtAlPRL2006}. 

Many of the previous approaches, however, operate on lattice geometries that are too simplified to mimic the aforementioned complexity of neural circuitry\cite{NowrouziNezhadEtAlPHYSICAA2025}. For one, neural connections are inherently directed and weighted, precluding the use of undirected regular networks commonly used in sandpile model implementations. More importantly, brain networks are known to exhibit hierarchical structures with modular effective functioning units\cite{SpornsNETWORKS2016}. Kaiser and Hilgetag\cite{KaiserFRONTIERS2010} utilized this fact to create hierarchical modular networks (HMNs) with small-world characteristics embedded in 2d substrates that represent brain connectivity on a large/mesoscopic scale, where the nodes represent cortical columns rather than individual neurons. They used this model to investigate topological effects on limited-sustained activity. \'Odor et. al.\cite{OdorEtAlSCIREP2015} used a variant of this and showed that 
Griffiths Phases (GP)\cite{GriffithsPRL1969} may emerge by a simple threshold model dynamics. 
The fact that pure topological inhomogeneity may result in GP in finite-dimensional networks\cite{MunozEtAlPRL2010}, or in Griffiths effects\cite{CotaEtAlPRE2016} in infinite dimensions, was confirmed by brain experiments\cite{PalmaEtAlPNAS2013}. GP is
suggested be a good candidate mechanism for explaining working memory in the brain\cite{JohnsonPLOSONE2013}. Additionally, GP provides an alternative mechanistic explanation for dynamical critical-like behavior observed without fine-tuning, consistent with the criticality hypothesis\cite{Moretti2013,OdorEtAlSCIREP2015,OdorPRE2016,OdorPRE.99.012113}. 

In addition to the use of more realistic network architectures, the models should also incorporate learning, the strengthening of network links through correlated neuronal activity and the corresponding weakening of unused pathways. Learning in critical state brain models was patterned after neuroscience modeling of empirical connectomes\cite{DeArcangelisHerrmannPNAS2010}. However, in these early approaches, the role of random weakening mechanisms in learning and system criticality has not yet been studied in detail. Recent results on connectome database analysis\cite{CirunayEtAlPRR2025} and magnetoencephalography and tractography suggest that this is the driving mechanism for the complexity of real connectomes\cite{AngiolelliEtAl2024}.

To address these research gaps, we revisit here the avalanche perspective on cortical dynamics and introduce refinements to the sandpile-based model of neural criticality. First, we use graph-dimension-controllable HMN structures for implementation, which are denser than the commonly used random networks in literature and are deemed to be more representative of the actual structure and dynamics of the brain\cite{OdorEtAlSCIREP2015}. Secondly, while keeping the avalanche dynamics similar to the sandpile rules, we introduce different rates of activation and inhibition mechanisms in the Hebbian learning. In particular, the rate of new link creation is deemed to be instantaneous for all sites reaching criticality over a single avalanche cycle; on the other hand, random weakenings of unused pathways are gradual, following an exponential decay over time. By modifying the proposed model's update rules to incorporate Hebbian learning, we show by extensive simulations that it can generate global PL weight distributions and LN ones for the degrees and node strengths. We provide finite-size scaling collapses and provide numerical evidence for this. We also investigate the effect of initial conditions by running simulations on HMN networks with different graph dimensions. Finally, to test the validity of the model, we provide comparisons with the actual edge weight distributions found in empirical connectome data.

\section*{Results} 
\label{sec:results}

In the following, we present the results of extensive simulations that run the locally conservative sandpile model with Hebbian learning on hierarchical modular networks (HMNs) where each node represents a neuron. For the baseline HMNs used, the number of nodes $N = 4^{l_{\max}}$, are varied, with $l_{\max} = \{5, 6, 7, 8\}$. The edges are weighted and directed, simulating synaptic connectivity. The average edge count per node, $\langle k_0 \rangle$, is the total number of edges $E_0$ of the initial HMN divided by the number of nodes $N$. We tested for $\langle k_0 \rangle \simeq \{7.6, 11.8, 13.9\}$, 
and most of the results presented are for $\langle k_0 \rangle = 11.8$ [see \blue{ Supplementary Information: Effect of Initial Link Density} for details]. Finally, we also looked at the effect of long-range connectivity: Starting from a single connected two-dimensional base lattice, long-range links are added among levels of the HMN whose probabilities of connectivity decay with Euclidean distance via $p(R) \sim \langle k_0 \rangle R^{-s}$; here, we considered cases where $s=3$ and $s=4$ [see \blue{ Methods: Hierarchical Modular Networks (HMNs)}]. Dynamical activity is then introduced into the baseline HMN through a locally conservative sandpile model [\blue{ Methods: Sandpile Model Implementation}], while network evolution is imposed through correlated link creation and random link weakening/removal [\blue{ Methods: Hebbian Learning}]. The state of the system is analyzed at the instant when the network achieves maximal connectivity [\blue{ Supplementary Information: Network Evolution}]. Analysis of the resulting avalanches and the network connectivity reveal strong indications of critical behavior, reminiscent of those found in actual connectomes.

\subsection*{Avalanche size distributions}
\label{ssec:avalanchesize}

\begin{figure}[h!]
    \centering
    \includegraphics[width=\columnwidth]{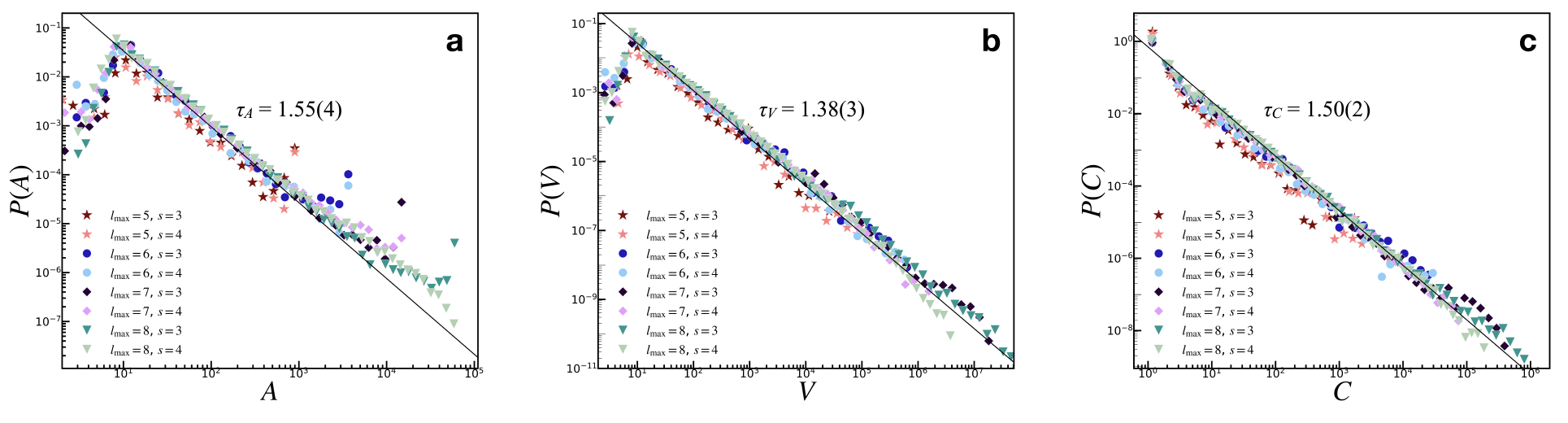}
    \caption{\label{fig:avalanchesize} \textit{Avalanche size distributions}. (a) The area distributions show powerlaw tails that follow $P(A) \sim A^{-\tau_A}$, where $\tau_A = 1.55(4)$. The tails of the distribution show increased occurrences of $A = N$, making the tails significantly deviate from the initial decaying trend. (b) The activation distributions follow $P(V) \sim V^{-\tau_V}$, where $\tau_V = 1.38(3)$. (c) The toppled sites count distributions follow $P(C) \sim C^{-\tau_C}$, where $\tau_C = 1.50(2)$ consistent with the mean field universality scaling and electrode experimental results in the human brain\cite{BeggsJNEURO2003}. Initial HMN graph parameters are displayed by the legends.}
\end{figure}

The first indication of critical behavior is the emergence of PL distributions of characteristic avalanche sizes similar to those obtained for the non-adaptive SOC models like the original sandpile model\cite{BakEtAlPRL1987} and other models with static rules in undirected lattice geometries\cite{ZhangPRL1989,DharPRL1990}. For the specific case of neuronal criticality, the PL exponents of $3/2$ obtained from experiments\cite{Beggs2022} are recovered ``for the high-dimensional, directed mean-field conserved sandpile universality class"\cite{DharRamaswamyPRL1989,ZachariouEtAlPLOSONE2015} with the upper critical dimension $d_c=3$. Here, we report robust PL distributions that span several orders of magnitude for various measures of avalanche sizes, one of which manifests the $3/2$ exponent of neuronal criticality.

In the implementation used here, individual sites contain continuous-valued ``grains'', which, for simplicity, reach a threshold value at unity. Continuous driving of the grid at random node sites results in avalanches when the threshold value is reached or exceeded at random node locations. We measure the extent of the avalanche in the network of nodes using three metrics. One is the simple count of the nodes that changed state during an avalanche event; in keeping with the literature on grid implementations, we call it the \textit{area}, $A$. Another is the count of the actual number of individual toppling and receiving events, in which nodes can be counted more than once based on their repeated participation in the entire avalanche; we call this the \textit{activation}, $V$, consistent with other previous works\cite{BatacEtAlNPG2017}. Finally, we also investigate the \textit{toppled sites count} $C$, which is the number of sites that exceeded their threshold states and toppled, redistributing the ``grains'' to their direct outward connections. This metric is inspired by the actual measurements done in neural connections, which only account for the total number of active nodes within an avalanche including multiple activations of the same nodes\cite{HeineyEtAlFRONTIERS2022, PasqualeNEUROSCIENCE2008}.  The $A$, $V$, and $C$ measures of avalanche magnitudes exhibit PL tails, indicating that they are just different characterizations of the same critical behavior operating within the network [see \blue{ Supplementary Information: Scaling Behavior of Avalanche Metrics} for details].

Interestingly, despite the highly dynamic nature of the rules, specifically those that govern the network evolution, we found very robust PL tails for areas, $P(A)$, activations, $P(V)$, and toppled sites, $P(C)$.  Fig.~\ref{fig:avalanchesize}(a) show the tails of the area distributions that follow a linear trend on a double logarithmic scale: starting from the modal value $\hat{A} = 10$, the tails obey a PL distribution $P(A) \sim A^{-\tau_A}$, where $\tau_A = 1.55(4)$. It should be noted that the tails of $P(A)$ show spikes due to the repeated occurrences of avalanches of size $A = N$ (i.e. spreading to all nodes), especially for very many iterations (an iteration refers to one instance of dropping a grain of sand), when the network becomes very densely connected. This highlights the limitations of the measure $A$ in capturing the extent of the activity in the network: the network of nodes can manifest spatiotemporal activities several orders of magnitude greater than the node count $N$, which $A$ cannot measure.

As such, the dynamical features of the network are better captured by the activation $V$, which is the spatiotemporal measure of each node's participation (both as a receiving or a toppling site) in the avalanche. The distribution of site activations, Fig.~\ref{fig:avalanchesize}(b), follow PL tails $P(V) \sim V^{-\tau_V}$, with $\tau_V = 1.38(3)$ , starting from the mode $\hat{V} = 10$. Unlike the area $A$, the activation $V$ extends beyond the physical limit of the number of nodes in the network, capturing individual toppling and redistribution of ``grains'' within the network. It is worth noting that the tails of the $P(V)$ distributions show a tapering at the tails, indicating the expected finite-size limitations on the activity of the network. The decrease in the powerlaw slope for $P(V)$ can be explained by the increased incidence of larger avalanches caused by the reactivations of ancestor nodes\cite{BeggsJNEURO2003}. 

Finally, the distribution of toppled site counts, Fig.~\ref{fig:avalanchesize}(c), show PL tails that obey $P(C) \sim C^{-\tau_C}$, with $\tau_C = 1.50(2)$. By counting only the sites that toppled during an avalanche event, regardless of whether they are initially directly connected or not, the model recovers similar statistics as those obtained from the local field potential measurements of electrodes on brain slices, replicating the well-known critical PL exponents of 3/2 for the neuronal avalanches as reported by Beggs and Plenz\cite{BeggsJNEURO2003} and in other brain experiments\cite{PlenzEtAlFRONTIERS2021}. Also, this avalanche universality occurs in directed models of firms\cite{ZachariouEtAlPLOSONE2015} among others. 

It is important to note that the obtained PL distributions are not sensitive to the sizes of the original HMN, $N = 4^{l_{\max}}$, for the broad range of $l_{\max} = \{5, 6, 7, 8\}$ considered. Furthermore, there is very little effect  of the other control parameters of the HMN. This suggests a mean-field-type of critical behavior, despite the variability of the network architecture. An explanation for this could be related to the rapid long-range interactions generated by the avalanches. We will show further evidence of scaling behavior via finite-size scaling.

\subsection*{Edge weight distributions}
\label{ssec:edgeweightdist}

\begin{figure}[h!]
    \centering
    \includegraphics[width=0.8\columnwidth]{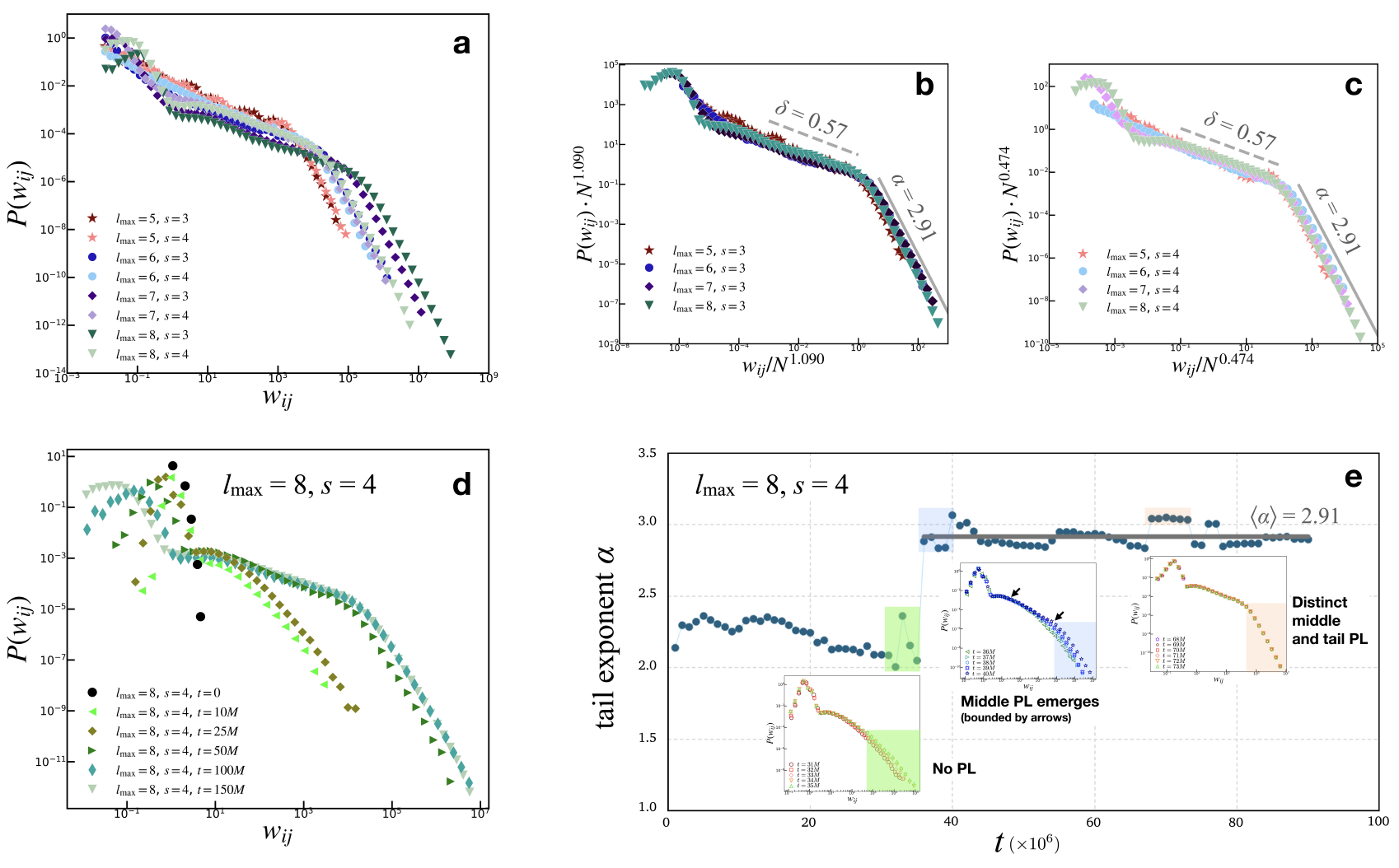}
    \caption{\label{fig:edgeweight} \textit{Global edge weight distributions.} [\textit{Top panels:} Results for various $l_{\max}$ and $s$ values for $\langle k_0 \rangle = 11.8$; \textit{Bottom panels:} Time evolution of the representative case $l_{\max} = 8$, $s=4$, $\langle k_0 \rangle = 11.8$.] (a) The $P(w_{ij})$ show two PL regimes that are better observed upon finite-size scaling: (b) the $s=3$ distributions collapse into a universal curve upon rescaling by $N^{1.09}$; and (c) for the $s=4$ distributions, $N^{0.474}$. For both (b) and (c), the rescaled curves show an intermediate regime with a gentle PL trend $P(w_ij) \sim w_{ij}^{-\delta}$, where $\delta = 0.57(4)$, and a tail with steeper PL, $P(w_ij) \sim w_{ij}^{-\alpha}$, where $\alpha = 2.91(5)$. (d) For the case of $l_{\max} = 8$, $s=4$, representative snapshots of the network for various instances of time show the gradual evolution of the $P(w_{ij})$, particularly the evolution of the PL regimes. (e) A detailed examination of the first 90 million iterations show the strong approach towards steeper tails $P(w_{ij}) \sim w_{ij}^{-\alpha}$, with $\langle \alpha \rangle = 2.91$; the PL trend with this exponent is shown by the gray guides to the eye in (b) and (c).
    }
\end{figure}

Unlike previous sandpile-based implementations that preserve the structure of the grid, the instantaneous link creation during avalanche events and the gradual decay of randomly selected links during stasis times (i.e. time instances with no avalanche events) produces a dynamic network that is deemed to better represent the interacting functional brain units. The state of the network at its peak connectivity, that is, when it has reached the maximum number of links, can be described by the distribution of edge weights $w_{ij}$, displayed in Fig.~\ref{fig:edgeweight}. 

\begin{figure}[h!]
    \centering
    \includegraphics[width=0.8\columnwidth]{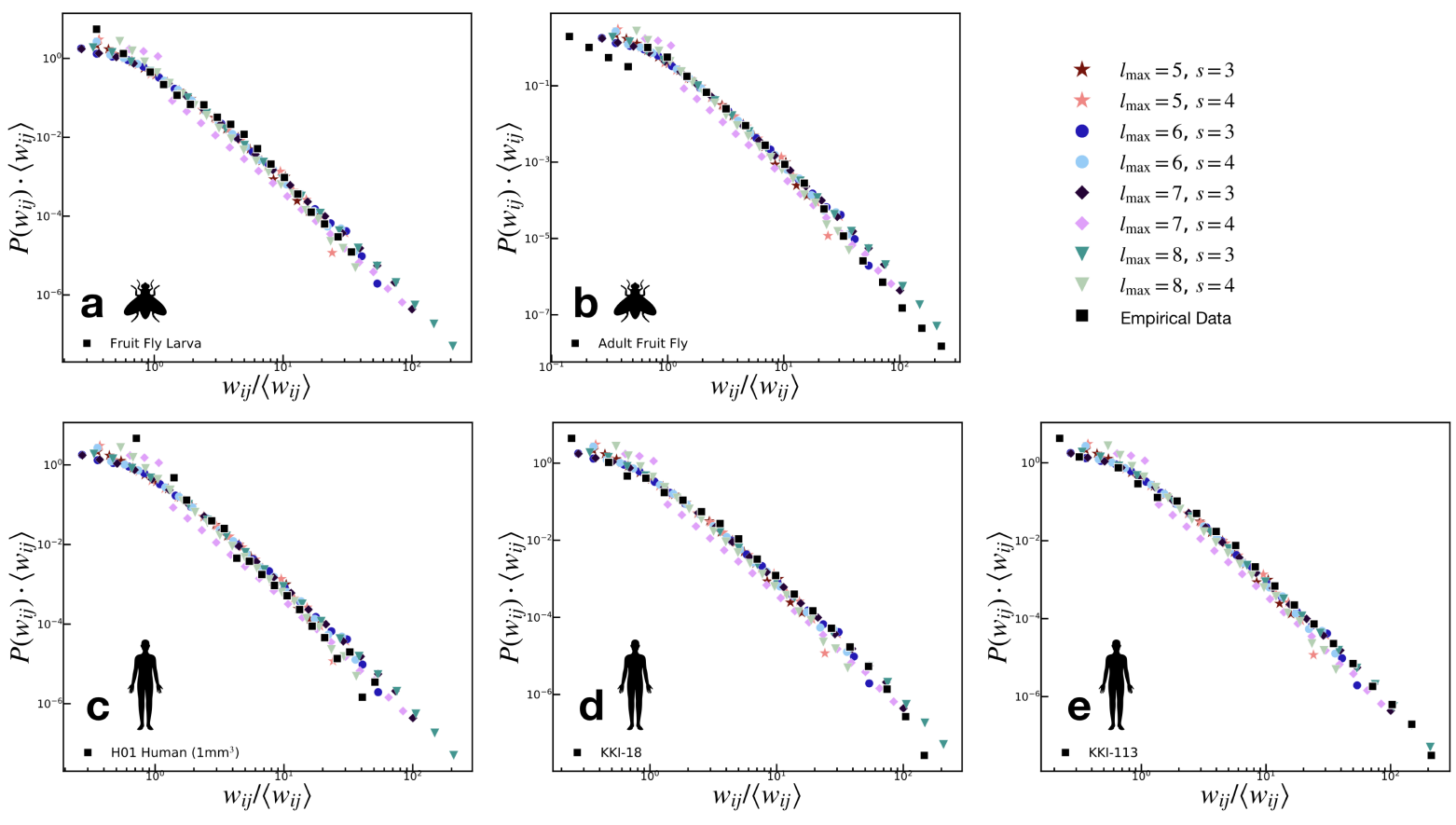}
    \caption{\label{fig:edgeweightmodeldata} \textit{Edge weight distributions: Model comparisons with data.} The model captures the distribution of edge weights from connectome data (arranged here in increasing levels of complexity): (a) fruit fly (larva) (b) fruit fly (adult) (c) H01 Human 1mm$^3$ (d) human KKI-18 and (e) human KKI-113.}
\end{figure}

For a sufficiently large number of iterations and before the network becomes fragmented, Fig.~\ref{fig:edgeweight}(a) shows the general trends in the distribution of $w_{ij}$ for $l_{\max} = \{5,6,7,8\}$ and $s = \{3,4\}$, of the original HMNs with an average edge count per node $\langle k_0 \rangle = 11.8$. The regime of the smallest edge weights, appearing as unimodal bumps for the smallest $w_{ij}$, is deemed to be an artifact of the original HMN construction. In particular, it is worth noting that these peaks are close to the tolerance value $w_{tol} = 10^{-2}$, which is the threshold for pruning links out of the network. These features of the distributions are changed, if not completely lost, upon the temporal evolution of the network. What follows is a ``middle'' region, which, especially for the largest baseline HMNs, show a slow PL decay. Above a certain knee value of $w_{ij}$, the distributions cross over to steep PL tails, with a common decay trend. These regimes, unlike those for smaller $w_{ij}$, are robust and continue to manifest for longer iteration times. 

In Fig.~\ref{fig:edgeweight}(b) and (c), the results for $s=3$ and $s=4$, respectively, show data collapses upon finite-size scaling (FSS). In Fig.~\ref{fig:edgeweight}(b), the $s=3$ data shows data collapse upon rescaling by $N^{1.090}$; on the other hand, in Fig.~\ref{fig:edgeweight}(c), the $s=4$ data shows the same collapse upon rescaling by $N^{0.474}$. The different FSS exponents suggest that the $s=3$ and $s=4$ weight distributions belong to different finite-size scalings. Note, that there is no indication for different avalanche size scaling behaviors in these cases. The case of $s=3$ in Fig.~\ref{fig:edgeweight}(b) shows FSS with $N^{1.090} \sim N$, corresponding to 
$d \simeq 4.7$ [see dimensions in Fig. \ref{fig:hmn}(b) under \blue{Methods: Heirarchical Modular Networks (HMNs)}]; on the other hand, in 
Fig.~\ref{fig:edgeweight}(c), the $s=4$ case with FSS factor of $N^{0.474}$ is related to $d \simeq 2.7$ topological dimension version of the graph.

In both cases, however, the strong data collapse further highlight the gentle middle PL and the sharp tail PL. Fig.~\ref{fig:edgeweight}(b) and (c) shows the best-fit PL exponents for the middle regime [dashed gray lines] and the tails [thick gray lines]. The gentle PL decay in the middle regions follows $P(w_{ij}) \sim w_{ij}^{-\delta}$, with $\delta = 0.57(4)$. The tails collapse to a PL $P(w_{ij}) \sim w_{ij}^{-\alpha}$, characterized by $\alpha = 2.91(5)$. To further illustrate the emergence of these regimes, Fig.~\ref{fig:edgeweight}(d) shows the representative case for one of the largest baseline HMNs considered, for $l_{\max} = 8$, $s=4$, and $\langle k_0 \rangle = 11.8$, tracked for various time snapshots. From the original HMN with an exponential distribution of edge weights, the $P(w_{ij})$ gradually evolve to produce the clear distinction between the two PL regimes. In particular, from around $t=80M$ iterations, the middle and tail PL regions of $P(w_{ij})$ begin to overlap, until the point of maximum connectivity at around $t = 174M$; in Fig.~\ref{fig:edgeweight}(d), the case of $t=100M$ and $t=150M$ are shown.

The distinction between the middle and tail regimes of $P(w_{ij})$ is further investigated by statistical tests for PL behavior. Using the largest network $l_{\max} = 8$ for $s=4$, the snapshots of the network are obtained for every 1 million iterations and their corresponding $P(w_{ij})$ are obtained. For each of these time snapshots, the last two orders of magnitude of the set of $w_{ij}$ are taken and tested for PL trends using the \texttt{powerlaw} Python package\cite{AlstottEtAlPLOSONE2014} that is based on the Kolmogorov-Smirnoff statistics\cite{ClausetEtAlSIAM2009}. The use of the last two orders of magnitude of the data makes it a good approximation for the tail regions. The algorithm tests for $P(w_{ij}) \sim w_{ij}^{-\alpha}$, where $\alpha$ is computed along with the minimum value of $w_{ij}$ that best recovers the PL fit. In Fig.~\ref{fig:edgeweight}(e), the computed $\gamma_2$ of the tails are plotted with the time snapshot $t$, showing a sharp transition at around 35 million iterations. The insets of Fig.~\ref{fig:edgeweight}(e) show the actual $P(w_{ij})$ plots, along with the chosen regions of PL from the algorithm.

For iteration times $t < 35M$, the distributions $P(w_{ij})$ exhibit tails showing no distinction between the two PL regimes and are thus qualitatively not best represented by PL. In the green shaded region of Fig.~\ref{fig:edgeweight}(e), the insets show the relative range of values chosen by the \texttt{powerlaw} algorithm to be the best approximation of the PL regimes. The lack of ``true'' PL regimes results in low computed $\alpha$ values for these runs. However, shortly after the $t=35M$ iterations, the middle region with a gentle PL slope begins to emerge. At this point in the learning, the short ranged correlations of the base lattice build up, resulting in the middle PL and  a crossover to the tail PL scaling behavior. As such, the \texttt{powerlaw} algorithm determined the ``true'' tails, resulting in the jump to the larger $\alpha$ values from this point onwards. The blue shaded region and the corresponding insets show the range of PL considered by the algorithm, which corresponds to the region beyond the knee. Finally, for the orange shaded region in Fig.~\ref{fig:edgeweight}(e), the tails (and, in turn, the middle region) are now well-defined, resulting in the continuous recovery of large $\alpha$. These regimes are the result of the gradual emergence of long-range correlated weight structured networks via learning. From this point onwards, the tail PLs show $\alpha$ values close to 3; the average of the runs from $t=36M$ to $t=90M$ is $\langle \alpha \rangle \approx 2.91$, which is also shown as the guides to the eye for Fig.~\ref{fig:edgeweight}(b) and (c).



Extensive and long-time simulations point to the strong convergence towards the $\alpha$ PLs at the tails. Interestingly, the values of $\alpha$ obtained by the model are close to the PL tails of the empirical connectome data. The top panels of Fig.~\ref{fig:edgeweightmodeldata} presents comparisons of the PL tails edge weight distributions of the model (normalized with their corresponding mean values) with those measured from connectome datasets of fruit fly [larva, Fig.~\ref{fig:edgeweightmodeldata}(a); and adult, Fig.~\ref{fig:edgeweightmodeldata}(b)] and humans [H01 1 mm$^3$, Fig.~\ref{fig:edgeweightmodeldata}(c); KKI-18, Fig.~\ref{fig:edgeweightmodeldata}(d); and KKI-113, Fig.~\ref{fig:edgeweightmodeldata}(e)], also rescaled with their respective average values. The model-generated distributions, regardless of network size and long-range connectivity, show remarkable correspondence with the empirical data, regardless of brain complexity.


\subsection*{Node degree and strength distributions} 
\label{ssec:strength}

\begin{figure}[h!]
   \centering
   \includegraphics[width=0.8\columnwidth]{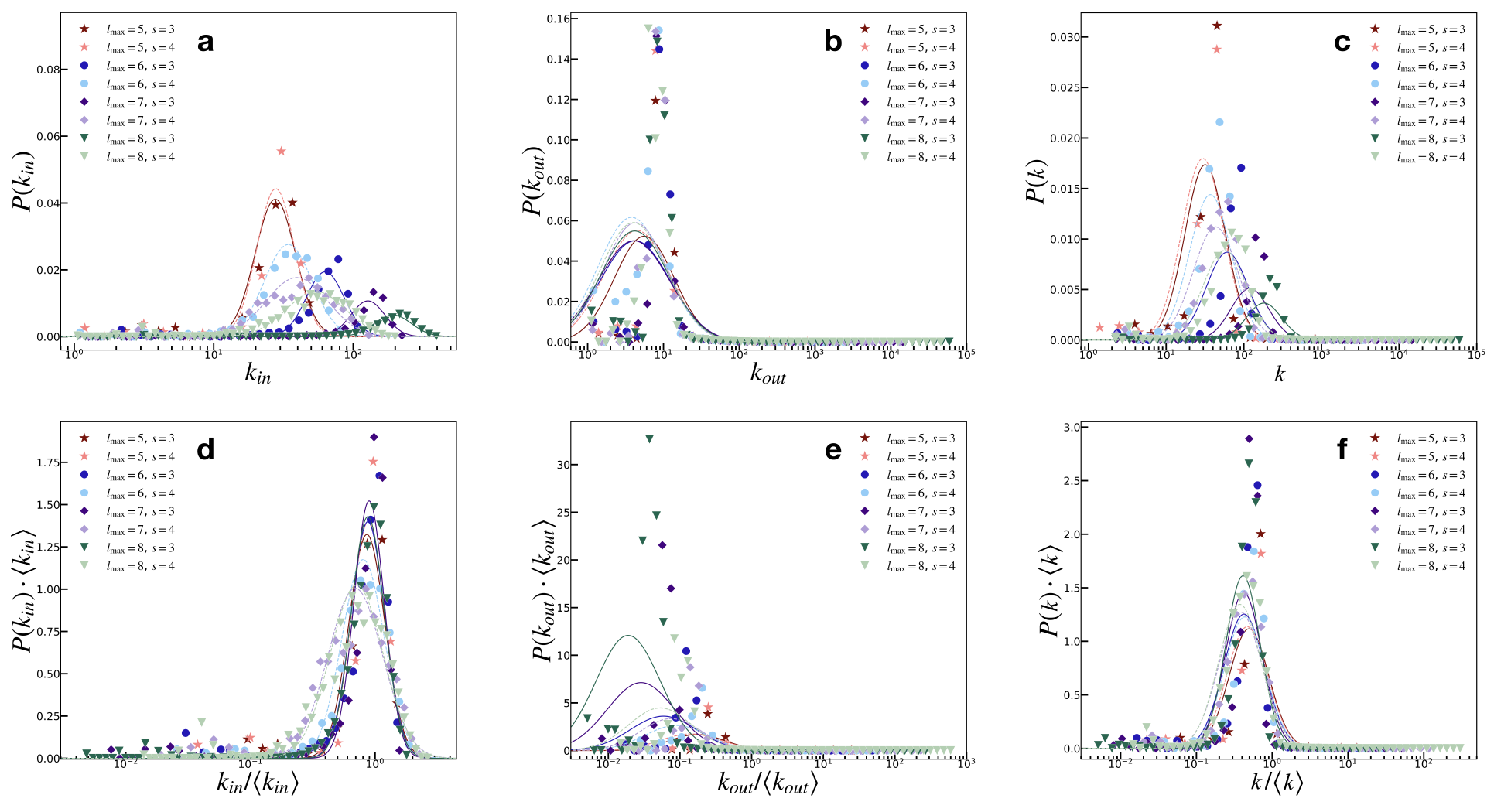}
   \caption{\label{fig:degree} \textit{Node degree distributions} [colored symbols are model results; LN fits are lines: solid lines for $s=3$, dashed lines for $s=4$]. The (a) $P(k_{in})$, (b) $P(k_{out})$, and (c) $P(k)$ are best fitted with lognormals, with characteristic (modal) values related to the original network size. (d)-(f) Rescaling with the mean results in a collapse into a common lognormal trend.}
\end{figure}

Fig.~\ref{fig:degree} shows the distribution of the in-degree $k_{in}$, out-degree $k_{out}$, and total $k = k_{in} + k_{out}$ degree for the case of maximum network connectivity [see \blue{ Supplementary Information: Network Evolution} for details]. In Fig.~\ref{fig:degree}(a)-(c), the node strength values are best fitted with lognormal (LN) distributions, here shown by the lines corresponding to the mean and standard deviation of the simulation results, with different shape and scale parameters resulting from different network sizes. A simple rescaling by the mean, however, shows a strong data collapse into a common lognormal trend, as shown in Fig.~\ref{fig:degree}(d)-(f). 

The recovery of LN node strength distributions strongly agrees with recent empirical observations on actual connectome data\cite{CirunayEtAlPRR2025}. While other authors empirically describe only the tails of the node strength distributions, approximating them as decaying PLs\cite{LynnEtAlNATUREPHYSICS2024}, there are good reasons to believe that the LN distributions obtained by the model is also the one manifested by real neuronal networks. In particular, the intrinsic restrictions imposed by basic spatial and metabolic constraints\cite{SpornsNETWORKS2016} on the density and number of connections that may be maintained at any given node may make it seem difficult for geographically embedded networks to demonstrate a pure scale-free activity, resulting in LN distributions with a characteristic mode and decaying trends at the upper tail.

\begin{figure}[h!]
   \centering
   \includegraphics[width=0.8\columnwidth]{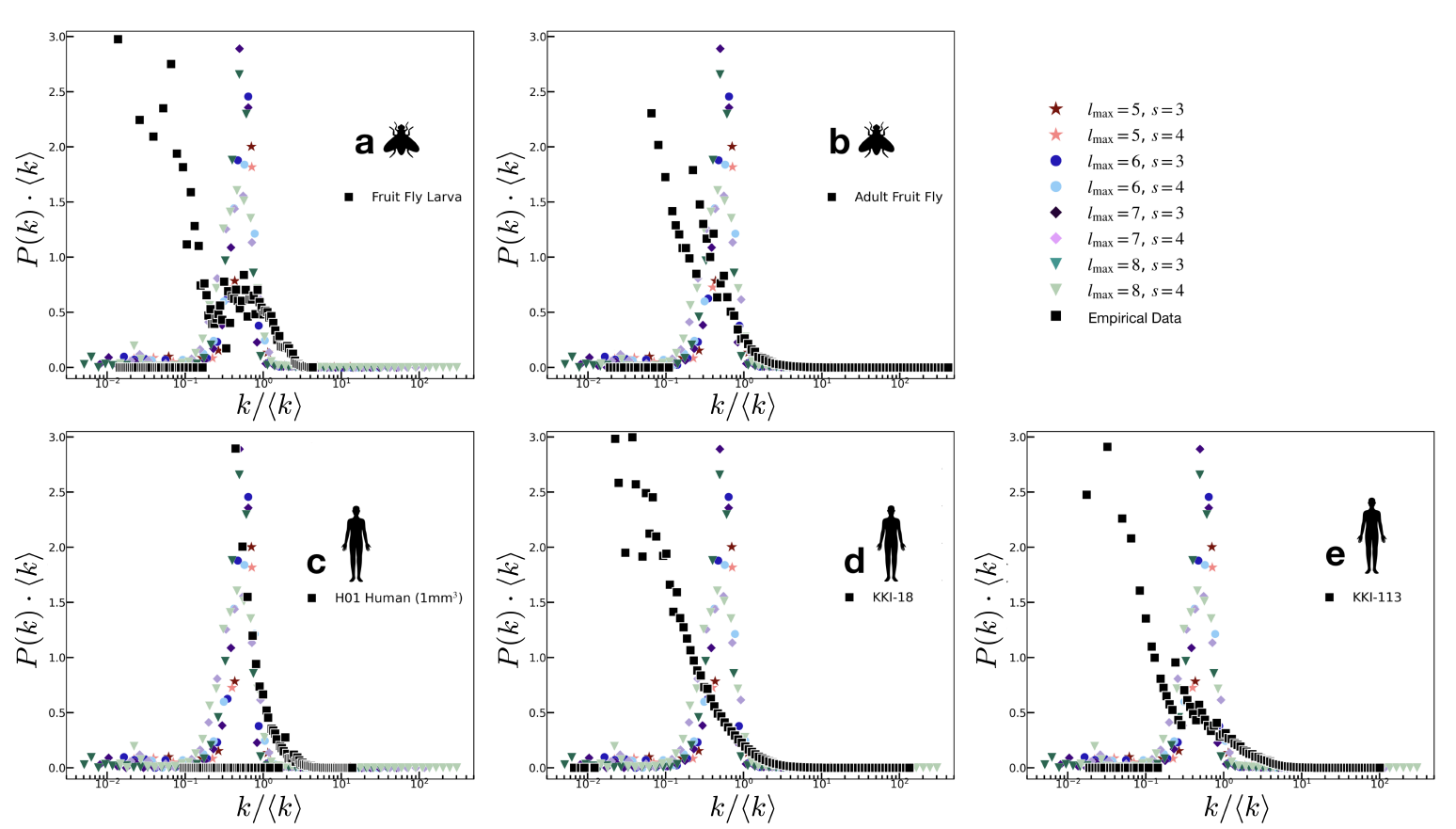}
   \caption{\label{fig:degree-datamodel} \textit{Node degree distributions: Model comparisons with data.} (a) Fruit fly larva (b) Adult fruit fly (c) H01 Human (1 mm$^3$) (d) KKI-18 (e) KKI-113. The unimodal behavior of the model recovers some features of the simpler and smaller connectomes: (a) fruit fly larva (b) adult fruit fly (c) H01 Human (1mm$^3$). All empirical results show a preponderance of small $k/\langle k \rangle$ values.}
\end{figure}

\begin{figure}[h!]
   \centering
   \includegraphics[width=\columnwidth]{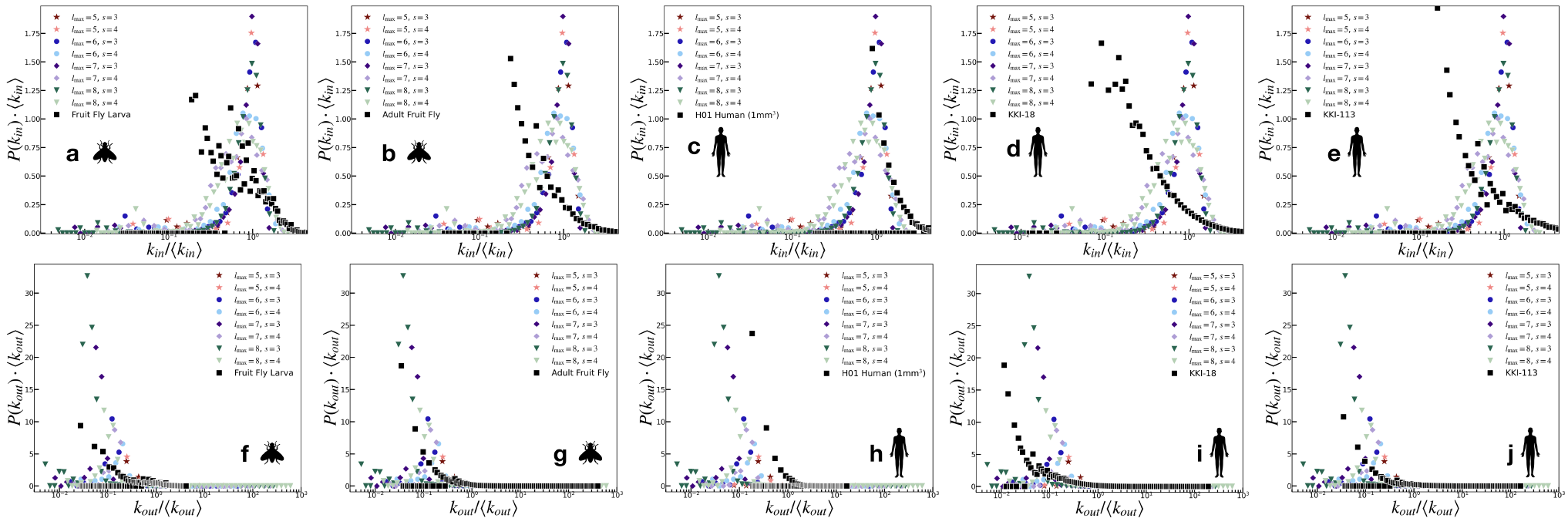}
   \caption{\label{fig:in_out-datamodel} \textit{Comparison of the in- and out- node strength distributions between data and model.} Top panels: The $P(k_{in})$ of the model fails to capture the high occurrences of smaller $k_{in}$ values of the data. Bottom panels: The tails of the  $P(k_{out})$ of the model and the data follow comparable decay trends.}
\end{figure}

Additionally, the local nature of the node degree that results in LN distributions stands in sharp contrast with the global distributions of edge weights, which follow PL trends. For this model, these statistical manifestations are the result of Hebbian learning that is strongly predicated upon the critical avalanche mechanisms. Larger avalanches produce more unstable sites over broader swaths of the network, producing more new outward directed links from the origin of the avalanche event. Furthermore, the strength of these new connections is proportional to the avalanche size (specifically, the activation), making these new connections less prone to the random pruning mechanisms. These aggressive mechanisms of strong link creations produce the broad tails of the strength distributions, particularly the $P(k_{out})$. It is therefore unsurprising that the presence of a characteristic modal values followed by heavy tails in the distributions of node strengths strongly aligns with previous works that show lognormal distributions in connectome data\cite{CirunayEtAlPRR2025}.

In Fig.~\ref{fig:degree-datamodel}, we present a comparison among the normalized model results with the same empirical connectome data sets presented in Fig.~\ref{fig:edgeweightmodeldata}. The strict unimodal distributions of the model replicates some features of the simpler connectome strength distributions, particularly those of the fruit fly larva [Fig.~\ref{fig:degree-datamodel}(a)], adult fruit fly [Fig.~\ref{fig:degree-datamodel}(b)], and the H01 human 1 mm$^3$ [Fig.~\ref{fig:degree-datamodel}(c)]. However, one clearly sees that the empirical data has a preponderance of smaller degree values; in fact, for more complex connectomes such as in the KKI-18 [Fig.~\ref{fig:degree-datamodel}(d)] and KKI-113 [Fig.~\ref{fig:degree-datamodel}(e)], these small $k/\langle k \rangle$ values dominate, such that the modes of the degree distributions are shifted left. The mismatch between the degree distributions of the complex connectomes is deemed to be a result of the lengthening of the tails of the empirical data: real neurons are more dynamic at creating new connections. The lengthening of the $P(k)$ tails of the data result in the mean being shifted rightward from the mode, which then shifts the mode of the data leftward from that of the model. 

In Fig.~\ref{fig:in_out-datamodel}, we present the comparison for the $P(k_{in})$ [Fig.~\ref{fig:in_out-datamodel}(a)-(e)] and $P(k_{out})$ [Fig.~\ref{fig:in_out-datamodel}(f)-(j)] between our model and the same set of empirical data as in Fig.~\ref{fig:degree-datamodel}. Similar to the observed behavior in Figure~\ref{fig:degree-datamodel}, the model shows comparable $P(k_{in})$ distributions as the smaller and simpler connectomes, particularly those of the fruit fly larva [Fig.~\ref{fig:in_out-datamodel}(a)], adult fruit fly [Fig.~\ref{fig:in_out-datamodel}(b)], and the human H01 1 mm$^3$ [Fig.~\ref{fig:in_out-datamodel}(c)]. The more complex connectomes KKI-18 [Fig.~\ref{fig:in_out-datamodel}(d)] and KKI-113 [Fig.~\ref{fig:in_out-datamodel}(e)] show higher probabilities of small $k_{in}$ values that is not captured by the unimodal trends of the model. On the other hand, we observe that the $P(k_{out})$ distributions from both empirical data and the model follow matching decay trends with tails that span several orders of magnitude as the mean. The mechanism of instantaneous link creation among firing nodes allows the model to recover the heavy-tailed distributions $P(k_{out})$.

\section*{Discussion}

In summary, we have presented here an implementation of a continuous sandpile model on a highly dynamic network topology. In previous works, sandpiles have been implemented on various network architectures\cite{ChenEtAlPHYSLETTA2008,GohEtAlPRL2003,PanEtAlPHYSICAA2007}, some of which are based on real-world\cite{ZachariouEtAlPLOSONE2015,WuEtAlPHYSICAA2024}. The novelty of the model presented here lies on the use of a baseline network that is inspired by the complex neural connectivity\cite{OdorEtAlSCIREP2015}, and the introduction of a network edge weight recalibrations, also inspired by the presumed dynamics of functional brain connectomes. 

One of the most striking results is the recovery of robust broad-tailed distributions of avalanche sizes [Fig.~\ref{fig:avalanchesize}], a key signature of critical behavior. By probing the area, activation, and toppled-site measures for a wide range of parameter values and network reconfigurations, we have shown robust PL tails across all network sizes and initial HMN wiring probabilities tested. In particular, the number of toppled sites during a single avalanche event shows PL distributions best-fitted with the 3/2 decaying exponent characteristic of local field potential electrode experiment neuronal avalanches\cite{BeggsJNEURO2003}. This result is deemed to be remarkable, given the highly dynamic nature of reconnections among the individual nodes in the network, which strongly affect the avalanching behavior of the system. The robustness of the avalanche-size distributions can be considered to be the manifestation of SOC mean-field universality, which occurs for $d>d_c=3$ for this model.

The model also reproduces the observed statistics of global edge weight distributions, whose tails can be fitted with decaying PLs with scaling exponents close to 3 in empirical data\cite{OdorPRE2016,OdorEtAlPRR2022,CirunayEtAlPRR2025}. The tails of the edge weight distributions are found to be robust across all the parameter sets tested [Fig.~\ref{fig:edgeweight}], and manifest remarkable comparisons with those obtained from actual connectomes [Fig.~\ref{fig:edgeweightmodeldata}]. 
An explanation for this can be given by the instantaneous avalanche propagation in mature networks close to maximal connectivity, which creates effective long-range interactions and smoothens out the spatio-temporal fluctuation effects. Under such conditions, a mean-field-like behavior emerges, explaining the emergence of the exponent 3 that closely matches that of the heuristic derivation for mean-field models in a previous work\cite{CirunayEtAlPRR2025}. Furthermore, this scaling, with some exponent variations within error margins, can also be the related to the heterogeneous HMN structure, which enhances rare-region effects in the modules. This means that even when the system is slightly super- (or sub-) critical, there can be arbitrarily large domains persisting in the opposite phase for long times, causing Griffiths effects, i.e. dynamical scale-free behavior in an extended parameter space. 
Our preliminary simulation results on homogeneous 4D lattices do not provide clear PLs as we see here for HMNs, strengthening the view that the network modularity plays an important role [see \blue{Robustness of Edge Weight Tail Distributions} for details]. 

Remarkably, the model also shows robust LN statistics for the node degree [Fig.~\ref{fig:degree}], which is also in agreement with the simpler connectome structures from empirical data [Fig.~\ref{fig:degree-datamodel}]. 

The model therefore simultaneously replicates the critical behavior of the brain from two distinct but related perspectives: (i) through the dynamic measures of brain activity\cite{BeggsJNEURO2003,HesseGrossFRONTSYSTNEUROSCI2014}, and (ii) through the static physical networks of neurons obtained from the connectome data\cite{OdorPRE2016,OdorEtAlPRR2022,CirunayEtAlPRR2025}. The latter is of particular importance; other discrete models only capture the criticality of neuronal avalanches, but not the statistics of the actual neuronal network structure. Here, the model we presented captures both the PL distributions of global edge weights and the LN distributions of local node degrees and sheds light on the adaptive structure-function relation in the brain\cite{AngiolelliEtAl2024}. 
The model does so by incorporating key details that are not considered in other works, namely, (i) the baseline HMN, and (ii) the adaptive strengthening [weakening] of nodal connections through correlated activity [inactivity].

In our view, the approach presented here provides a balanced modeling perspective. On one hand, critical behavior has been traditionally captured by the simplest models that self-organize and require no fine-tuning and specificity\cite{BakEtAlPRL1987}; but directly applying these approaches without proper context may lead to misleading or incorrect description of neuronal dynamics. On the other hand, the brain is perhaps one of the most complex and adaptive systems, whose associated functions are not yet completely attributed to structure and form; it is impossible to provide a complete description of the brain, as we are just starting to probe deeper into its physical intricacies. Our approach therefore has used the straightforward rules of discrete self-organizing models, while also incorporating the realistic conditions learned from recent physical investigations of the brain. The success of the model in recovering the signatures of neuronal activity and the physical network structure of the brain itself is predicated on two crucial components: (i) the retention of the criticality from the sandpile model; and (ii) the realistic depiction of learning on a hierarchical and modular network baseline that is inspired by the actual brain structure and dynamics.

The model may benefit from further details that mimic the realistic conditions in cortical dynamics. On the modeling side, additional factors such as local dissipation via ohmic losses, random fluctuations, and oscillatory mechanisms, among others, are not incorporated in the continuous sandpile used here. Moreover, subsequent implementations may benefit from incorporating some other features of neuronal dynamics; in particular, there could be reciprocity in link creation, which, when added to the model, could address the recovery of heavier tails of the degree distributions of more complex connectome empirical data [Fig.~\ref{fig:degree-datamodel}]. We believe that these attributes can easily be implemented for a more nuanced description. The rich dynamics and these hints of similarities with real systems further signify the emergence of complex behavior from simple rules, resulting in robust statistical regularities.

\section*{Methods}

\subsection*{Hierarchical Modular Networks (HMNs)}

To mimic the complex network connectivity of the brain in the modeling, we utilized a variant of HMNs, introduced by Kaiser and Hilgetag\cite{KaiserFRONTIERS2010} and modified by \'Odor et.al\cite{OdorEtAlSCIREP2015} to guarantee simple connectedness of nodes on the base level. This is constructed from nodes embedded on a square lattice with additional long links between randomly selected pairs, by an iterative rewiring algorithm of the following steps.

\begin{figure}[h!]
    \centering
    \includegraphics[width=0.7\columnwidth]{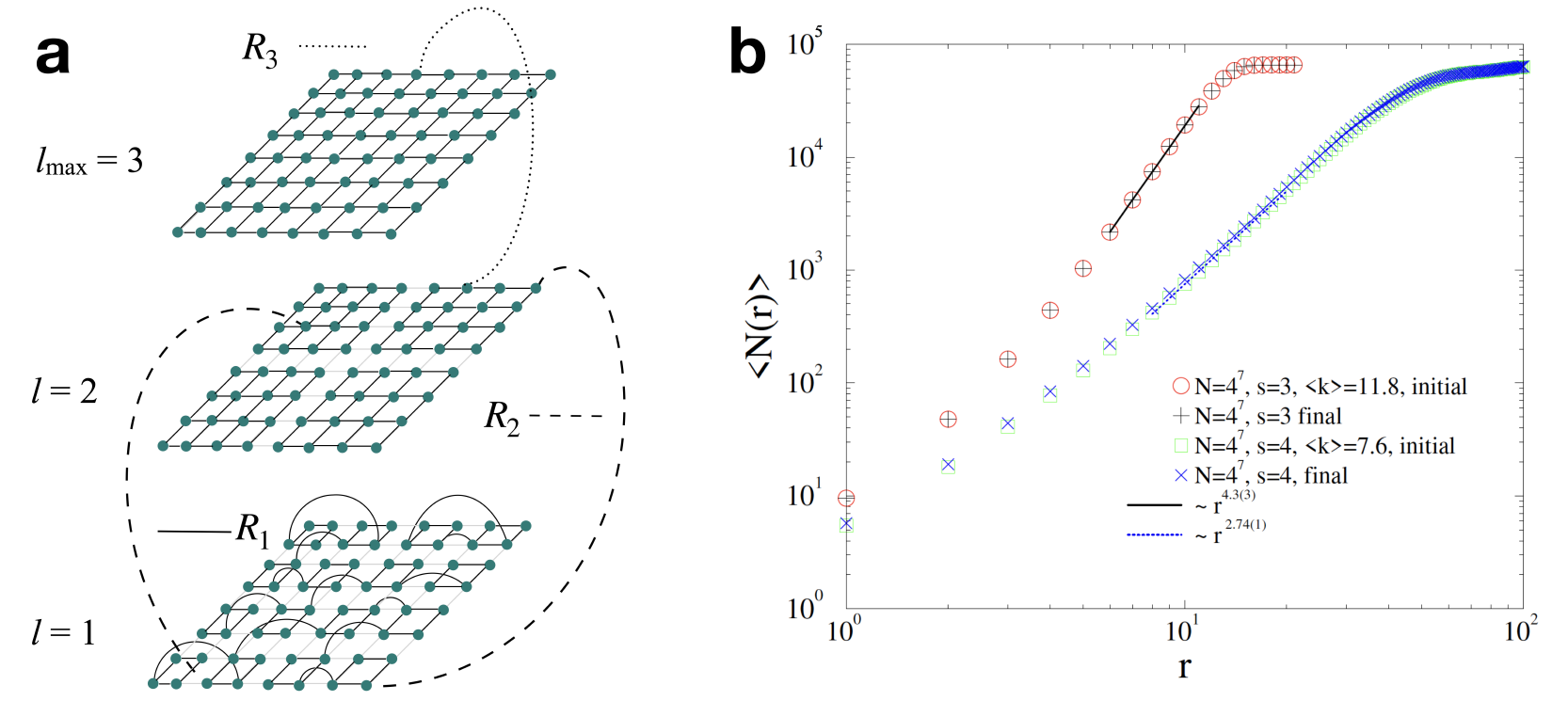}
    \caption{\label{fig:hmn} \textit{HMN construction and characterization.}  (a) Illustrating a two-dimensional HMN (HMN2d) with $l_{max} = 3$. Nodes at the bottom level $l=1$ are fully connected. Four bottom modules are grouped on the next hierarchical level to form one upper-level module and so on. $R_1$ (solid lines) denotes the randomly chosen connections among the bottom level; $R_2$ (dashed lines) and $R_3$ (dotted lines) denote the connection to the second level, and third level respectively. (b) Characterizing HMN: Average number of shortest path, chemical distances between nodes of $l_{\max}=7$ level HMNs, for $s=3$, $\langle k\rangle\sim 11.8$ and $s=4$, $\langle k\rangle\sim 7.6$ 
    before [hollow symbols] and after [crosses, plus signs] the learning. 
    Lines show powerlaw fits before saturation of the curves, describing the topological (graph) dimensions, defined by Eq.~(\ref{eq:dim}), similarly as in
    \cite{OdorEtAlSCIREP2015}.}
\end{figure}

To generate the HMN we define $l_{max}$ levels in the same set of $N = 4^{l_{max}}$ nodes. At every level, $4l$ modules are defined. Each module is split into four equal-sized modules on the next level, as if they were embedded in a regular, two-dimensional (2d) lattice. This is done because HMNs in 2d base lattices are closer to the real topology of cortical networks\cite{OdorEtAlSCIREP2015}. The probability $p_l$ that an edge connects two nodes follows $p_l \approx b2^{-sl}$ where $l$ is the greatest integer such that the two nodes are in the same module on level $l$ and $b$ is related to the average node degree via $b = \langle k_0 \rangle/2$. To illustrate, Fig.~\ref{fig:hmn}(a) shows a representative 3-level HMN2d hierarchical network construction. Here, the network is expanded to show how modules are defined on every level.

The addition of long-range graph links provides connections beyond the nearest-neighbor ones of the 2d substrate to mimic brain. Taking these into account, the HMN2d being utilized contains short links connecting nearest neighbors and long links, whose probability decays algebraically with Euclidean distance $R$,
\begin{equation}
\label{eqn: prob_R} p(R) \sim R^{-s}
\end{equation}

\noindent which makes it an instance of a Benjamini-Berger (BB) networks\cite{BenjaminiRANDSTRUC2001}. In this work, we consider a modified BB network in which the long links are added level by level, from top to bottom\cite{KaiserFRONTIERS2010}. The levels: $l = 1, \ldots, l_{max}$  are numbered from bottom to top. The size of domains, i.e., the number of nodes in a level, grows as $N_l= 4^l$ in the 4-module construction, related to a tiling of the two-dimensional base lattice. Due to the approximate distance-level relation, $R \sim 2^l$ , the long-link connection probability on level $l$ is:
\begin{equation}
\label{eqn: prob_R_HMN} p(R) = b \left( \frac{1}{2^s}\right)^{l}
\end{equation}

\noindent where $b$ is related to the average degree of the node. Here, we consider two-dimensional HMNs of different sizes $l_{max} = \{5, 6, 7, 8\}$, where, again, $N = 4^{l_{max}}$. We also used with different $s$ values, $s = \{3,4\}$ by providing single-connectedness on the base lattice. 


We define $r$ as the shortest path (chemical) distance and the average number of nodes within this distance 
$\langle N(r)\rangle$ is measured by the Breadth First Search algorithm~\cite{SihotangJINFO2020} emanating from all nodes of the graph. The topological (graph) dimension ($d$), can be defined as
\begin{equation}\label{eq:dim}
    \langle N(r)\rangle \propto r^{d}
\end{equation}

\noindent This is infinite for $s=3$ as $N\to\infty$, which appears to be $d = 4.3(3)$ in Fig.~\ref{fig:hmn}(b), in case of a finite graph with $l_{max}=8$ levels~\footnote{By increasing the size this effective dimension estimate would diverge.} In case of $s=4$ (for $N\to\infty$) one can obtain finite dimensions with $\langle k_0 \rangle$ initial degree dependent values.  In Fig.~\ref{fig:hmn}(b) one can fit $d = 2.74(1)$ for $s=4$, $\langle k_0 \rangle=7.6$,  which is below the upper critical dimension of this directed sandpile model ($d_c=3$).

We can observe that the learning does not seem to change the graph dimension (the shortest path lengths between nodes); instead, it only increases the average degree and the in/out strengths of nodes. This also means the lack of synaptogenesis in our model, which really happens in adult animals, only the synaptic weights increase.

One possible explanation for why the topological dimension of the HMNs does not change after imposing the model is the modular structure of the HMNs. Modularity aids in maintaining the overall structure of the network. During an avalanche, the span of the cascade may occur only within a module unless the avalanche is too big. As such, only the local connectivity is improved but does not dramatically change the topological properties of a modular network resulting to an unchanged dimension. In addition to this, the edge weight reinforcement rule includes a factor of $1/r$ which accounts for the realistic setting where the creation and maintenance of synapses in neural connectomes is accompanied by wiring cost\cite{CirunayEtAlPRR2025, AhnPHYSA2006, 
BullmoreNATUREREV2012} 
[see \blue{ Methods: Hebbian Learning} for details]; this appears to suppress the creation of very long links within the network.

\subsection*{Sandpile Model Implementation}

The cellular automata (CA) model of the sandpile is composed of a network of nodes, each representing a functional unit in the neural circuitry. Each node $i$ is characterized by its state $z_i$, a measure of its activity; in actual neurons, this can represent the action potential that spikes during firing. Because neuronal sections can only accommodate a finite potential, we set $z_i \in [0, z_{\max})$, where $z_{\max}$ is set to unity. The threshold state $z_{\max}$, when reached or exceeded by the node, triggers the relaxation of the site through the redistribution of stress to its direct neighbors. Finally, to also mimic the physical structure of the brain, the nodes are given physical site locations $(x_i, y_i)$ in a two-dimensional square-lattice grid. 

\begin{figure}[ht!]
    \centering
    \includegraphics[width=\columnwidth]{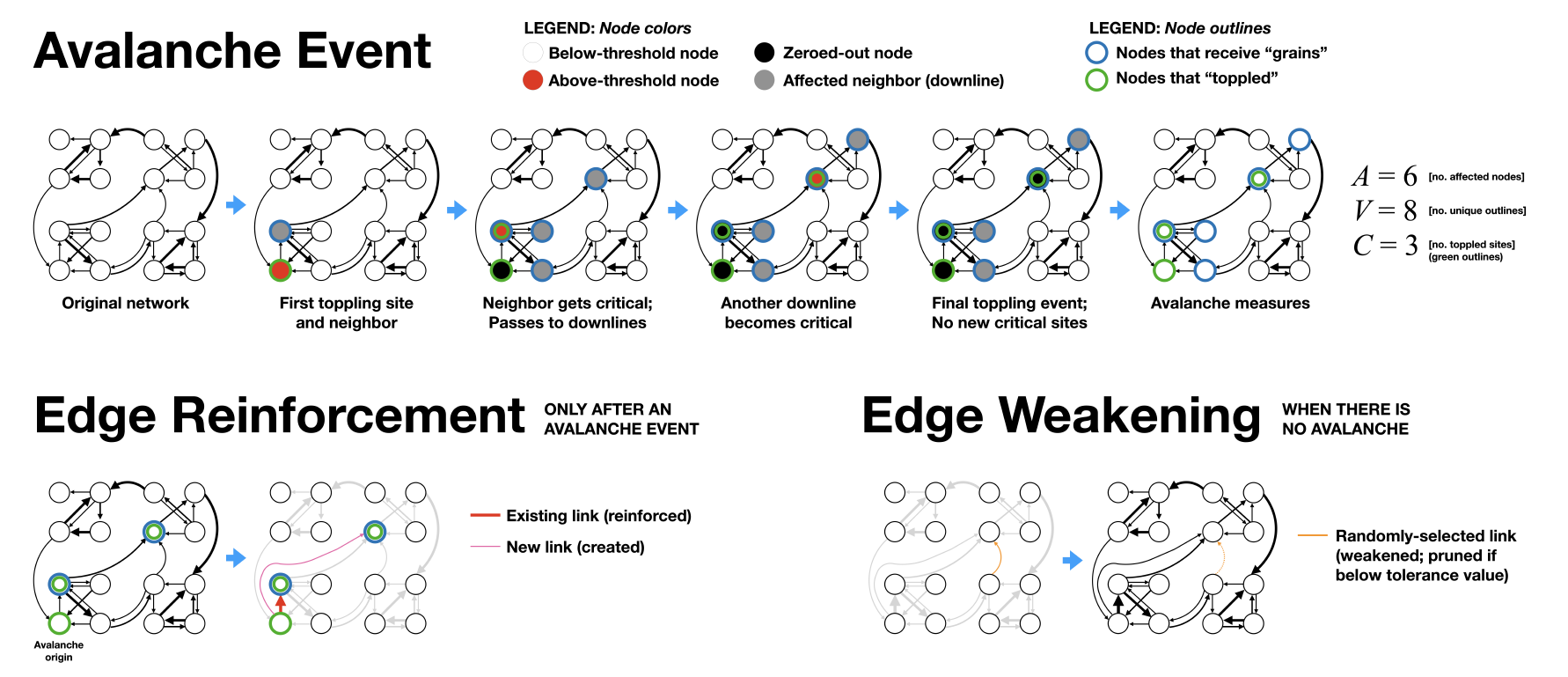}
    \caption{\label{fig:avalanchelearning}\textit{Avalanches and Hebbian learning.} The participation of the nodes in the original toppling event is measured as $A$, $V$, or $C$. During avalanche events, sites that toppled as a result of the original toppling event (at the avalanche origin) receives reinforced (or entirely new) connections based on Eqn.~\ref{eqn:enhance}. When there is no avalanche event, a random link is weakened (or entirely pruned) based on Eqn.~\ref{eqn:decay}.} 
\end{figure}

The network structure is introduced through a hierarchical modular network (HMN) [see previous section \blue{ Methods: Hierarchical Modular Networks (HMNs)}]. The directed connections between nodes $i$ (origin) and $j$ (destination) are represented by the edges $e_{ij}$, which is characterized by its weight $w_{ij}$, a proxy for the relative strength of synaptic connectivity between two brain sections. Mathematically, the HMN2d can now be represented as a graph $\mathcal{G} = \{\mathcal{V}, \mathcal{E}$\}, where the set of nodes $\mathcal{V} = \{i\}$ has $N$ elements, while $\mathcal{E} = \{e_{ij}\}$ contains $E$ number of connections; here, majority of the results are $\langle k_0 \rangle = E_0/N \simeq 11.8$ for the original HMN2d. Upon defining the network structure, the individual nodes $i$ now have a set of neighbors $n_i = \{j\}$, where $j$ are the out-degree connections 
or downlines of $i$. Whereas in simple grid implementations, the edges are bidirectional, uniformly weighted, and fixed to the discrete (usually, von Neumann) nearest neighbors in a grid, the HMN construction results in connections that are asymmetric $e_{{ij}} \in \mathcal{E}$ does not guarantee 
that $e_{ji} \in \mathcal{E}$ , i.e. $w_{ij} \ne w_{ji}$. 

In this context, it is also instructive to define the peripheral nodes of the network $\{i_p\}$. These nodes form the topological ``margins'' for ``overflow'' during avalanche events. In an actual grid setting, the periphery is the literal margins of the grid, whose states are zeroed out every iteration. Upon using HMNs as baseline networks, however, the notion of physical extremity is lost because of the structure of the network. Therefore, to be able to assign the nodes of egress, we used the betweenness centrality as a measure of periphery\cite{BatacCirunayPHYSICAA2022}. The states of the nodes with zero betweenness centrality are designated to be $\{i_p\}$ and zeroed out $z_i\to 0$ at every instant when they receive grains from the external driving or from the redistributions from their connections in the network. The egress of ``grains'' at the $\{i_p\}$ nodes ensure that, every time step, all the nodes in the network will ``relax'' to have finite and below-threshold values of the states.

During initialization, the nodes are given random states $z_i \in [0, z_{\max})$ drawn from a uniform distribution. Dynamic activity is represented by the introduction of a trigger to random nodes every time step, akin to the continuous reception of a background stimulus in actual neural activity. Here, the trigger strength $\Delta z = 10^{-4} z_{\max}$ increases the state $z_i$ of the randomly chosen node $i$ at an iteration time $t$, 
\begin{equation}
\label{eqn:externaltriggering} z_i^{t} \leftarrow z_i^{t-1} + \Delta z
\end{equation}

\noindent where the temporal index is placed at the superscript. Needless to say, repeated triggerings may result in the approach to the marginally stable state; such a site $i$ will be driven to $z_i^t = \bar{z} \geq z_{\max}$. When such an event occurs, time is frozen (i.e. no new $\Delta z$ is added to the system) while this above-threshold state is ``toppling,'' i.e. its above-threshold state $\bar{z}$ is being redistributed to its downline neighborhood in the network, $n_i = \{j\}$. Each downline neighbor receives a fraction of $\bar{z}$ that is proportional to the ratio of the individual edge weight $w_{ij}$ it receives from $i$ over the total out-degree edge weight of the neighborhood, $W_{n} = \sum w_{j}$:
\begin{equation}
\label{eqn:redistribution} z_{j} \leftarrow z_{j} + \frac{w_{ij}}{W_{n}} \bar{z} \textrm{; such that,} 
\end{equation}
\begin{equation}
\label{eqn:relaxation} z_i \leftarrow 0
\end{equation}

\noindent The redistribution and relaxation rules, Eqn.~(\ref{eqn:redistribution}) and (\ref{eqn:relaxation}) are repeated until all the nodes in the HMN have states below $z_{\max}$. Repeated redistributions affect portions of the network, resulting in one distinct avalanche event for the same instance of the initial triggering at $t$. Here, we consider two measures of avalanches used in traditional sandpile studies: the raw count of all the unique affected sites is called the area $A^t$ (the name being inspired by the extent of the avalanche in the usual 2D grid implementations); the number of multiple activations, counting multiple relaxations and/or receiving of redistributed states for all affected sites, is denoted by $V^t$. Furthermore, the individual toppled sites during a single triggering event is counted and labeled as $C$, corresponding to the measures of activated sites in actual experimental protocols\cite{BeggsJNEURO2003}. A representative example of an avalanche event and the corresponding avalanche measures is shown in Fig.~\ref{fig:avalanchelearning}.

\subsection*{Hebbian Learning}

During initialization, the starting HMN has discrete-valued i.e. $w_{ij} \in [0,...N]$ edge weights corresponding to the number of links between two nodes. To dynamically alter the network architecture, we added rules that strengthen and weaken the existing links based on measures of correlated activity and inactivity, respectively. These two mechanisms are guided by different mechanisms, and thus have different characteristic rates. On one hand, firing neurons tend to form stronger links within shorter time frames; on the other hand, less favored pathways tend to progressively decay in strength over extended periods of time. These foundational principles of Hebbian learning in the brain are simulated in the model through separate mechanisms of link creations and weakenings during avalanche and stasis periods, respectively.

After an avalanche event originating from $i$ at time $t$, all the other sites $k$ that reached $z_k \geq z_{\max}$ are tracked as candidate sites that will receive new connections $e_{ik}$ from $i$. The weight $w_{ik}$ of this new synaptic connection should reflect the mechanisms that enhance and hinder the connectivity of these functional units. Here, we impose that this weight be proportional to the avalanche measure, particularly the activation $V$; stronger connections are expected for multiple repetitions of correlated firings. Meanwhile, in a realistic setting, the connectivity is hindered by physical distance; here we use the simple distance $r_{ik} = \sqrt{(x_k - x_i)^2 + (y_k - y_i)^2}$ based on their locations on the square grid. With these assumptions, new links are created from $i$ to all $k$ nodes, with weights computed as
\begin{equation}
\label{eqn:enhance} w_{ik}^{t+1} \leftarrow w_{ik}^t + V^t / r_{ik}
\end{equation}

\noindent This rule instantaneously updates the neighborhood (downlines) of $i$ for the next time step $t+1$, with the nodes $k$ now joining the neighborhood $n_i = \{j\}$ and new connections $e_{ik}$ are added to $\mathcal{E}$. The ratio $V^t / r_{ik}$ is deemed to be the simplest representation of the effect of the avalanche and the separation distance on connection strength; increasing powers of $r_{ik}$ (say, $1/r^2$) significantly decreases the strength and spatial extent of new connectivity, especially for larger networks. 

What happens to the grid during periods of stasis? Here, we impose weakenings of random edges, albeit at a significantly slower rate; in reality, we don't expect neural connections to be severed instantaneously. The simplest assumption for the decay in edge weight is an exponential decay, wherein the subsequent (weakened) value is proportional to the current value. Therefore, for every time step with no avalanche event, a randomly chosen node $m$ and one of its neighbors $n$, originally connected by an edge $e_{mn}$ with weight $w_{mn}^t$ at time $t$, will have an updated connection weight, 

\begin{equation}
\label{eqn:decay} w_{mn}^{t+1} \leftarrow \beta w_{mn}^t 
\end{equation}

\noindent where, here, $\beta < 1$ is the decay rate. The repeated reinforcement [weakening] of the edges results in the edge weights being continuous valued. Additionally, when weakened down to a tolerance value $w_{tol}$, the edge $e_{mn}$ is completely severed, $w_{mn} = 0$. In some cases, for very long iteration times, this can lead to some nodes being isolated when all of their edges are removed. Here, we consider $w_{tol} = 0.01$, i.e. $1\%$ of a unit connectivity. For this $w_{tol}$ and all the starting HMN considered, we used $\beta = 0.99$; rapid deterioration of the network is observed for $\beta < 0.99$.

\section*{Supplementary Information} 

\subsection*{Network Evolution} 

\begin{figure}[h!]
    \centering \includegraphics[width=0.8\columnwidth]{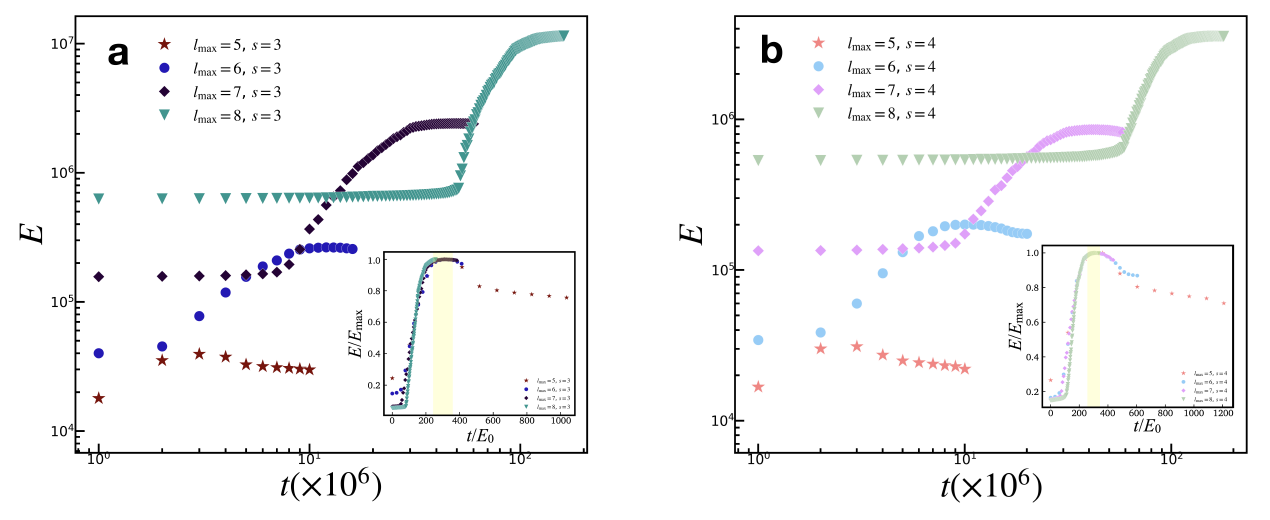}
    \caption{\label{fig:suppl-steadystate} \textit{Network time evolution.} Top panels: The total number of edges $E$ vs. time for (a) $s=3$ and (b) $s=4$. \textit{Insets}: When time $t$ is normalized with the number of links of the original HMN $E_0$, the plots show overlapping peaks that center at around $t/E_0 = 300$. The results presented in the paper used time snapshots of the network within the shaded region, $t/E_0 \in (250, 350)$.}
\end{figure}

Because of the relatively large number of parameters and the broad range of values considered, there should be a way to obtain a reasonable comparison among the results from different parameter sets. Thankfully, the model shows a remarkable collapse into a common temporal trend upon rescaling. For all the HMNs with different sizes $N^{l_{\max}}$, and for $s=3$ and $s=4$, the same trend in the eventual number of links is achieved when time is rescaled by $t/E_0$, where $E_0$ is the number of links in the original HMN. Fig.~\ref{fig:suppl-steadystate} shows the trends for all $l_{\max}$ and $s$ values.

The number of links within the network for the various instances of time as shown in Fig.~\ref{fig:suppl-steadystate}(a) and (b), presented here in double logarithmic scale for clarity, show an initial increasing trend, corresponding to the dominance of the link reinforcement/creation mechanisms. The shaded regions in the insets of Fig.~\ref{fig:suppl-steadystate}, corresponding to the values of around 250 to 350 of the ratio $t/E_0$, is deemed to be the case when the network has maximal connectivity. The actual instances of maximum network connectivity correspond to the following $t^\ast$ values: $t^\ast \approx 2$~million for $l_{\max} = 5, s=3$, and $4$~million for $l_{\max} = 5, s=4$; $12$~million for $l_{\max} = 6, s=3$, and $20$~million for $l_{\max} = 6, s=4$; $60$~million for $l_{\max} = 7, s=3$, and $80$~million for $l_{\max} = 7, s=4$; and $150$~million for $l_{\max} = 8, s=3$, and $170$~million for $l_{\max} = 8, s=4$. Unless otherwise specified, the results presented in the main text are obtained for network snapshots at $t^\ast$ or within the shaded regions in the insets of Fig.~\ref{fig:suppl-steadystate}.

Beyond this regime of maximal connectivity, the weakening mechanisms begin to manifest in the network, sometimes even removing the peripheral nodes. These disconnections result in the gradual decay in the number of links. The simulations are halted when all the peripheral nodes are removed from the system; in such a case, the ``grains'' would have no means for egress and the system can never achieve a relaxation. 

\subsection*{Effect of Initial Link Density} 

Starting with the same number of nodes $N$, the number of links $E_0$ of the baseline HMN is varied, producing different initial average degree values $\langle k_0 \rangle = E_0/N$. Here, we present some representative results for $\langle k_0 \rangle = 7.6$.

\begin{figure}[htbp!]
    \centering
    \includegraphics[width=\columnwidth]{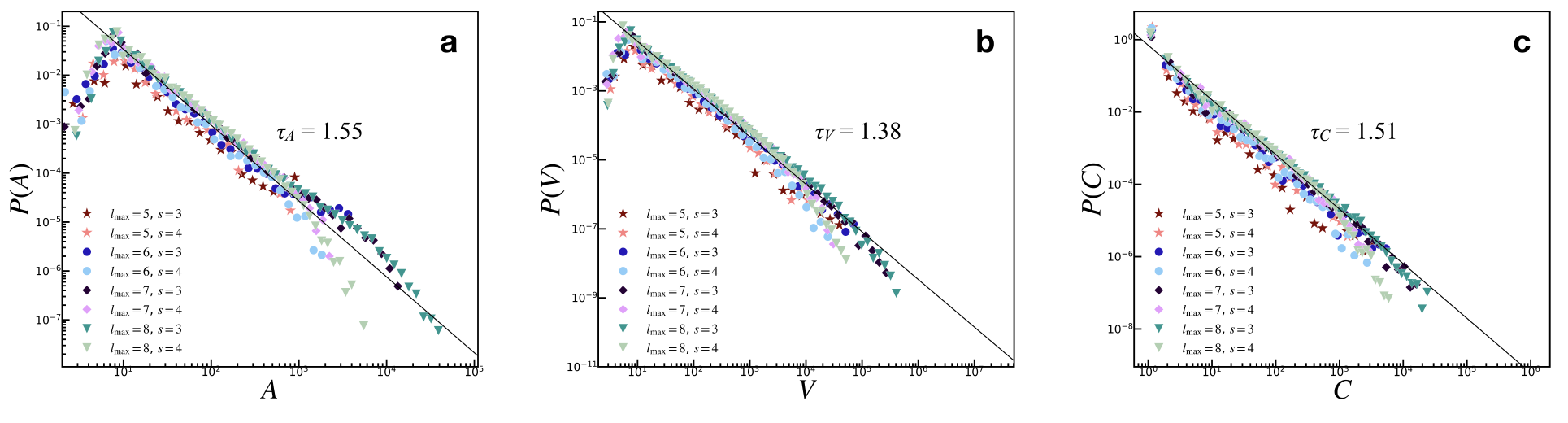}
    \caption{\label{fig:suppl-avc-e06n} \textit{Avalanche size distributions for $\langle k_0 \rangle = 7.6$.} The (a) $P(A)$, (b) $P(V)$, and $P(C)$ distributions follow similar trends as in Fig.~\ref{fig:avalanchesize} [for the denser $\langle k_0 \rangle = 11.8$ case], albeit with shorter tails.
    }
\end{figure}

Fig.~\ref{fig:suppl-avc-e06n} shows the distributions of the avalanche measures, (a) $P(A)$, (b) $P(V)$, and (c) $P(C)$ for $\langle k_0 \rangle = 7.6$. The axes scales are preserved for direct comparison with Fig.~\ref{fig:avalanchesize}, which are obtained for $\langle k_0 \rangle = 11.8$. The lower average number of links of the baseline HMN resulted in shorter tails for the avalanche size distributions, although the PL exponents remain almost the same. The insensitivity of the PL exponents to the initial HMN edge density is a clear indication of critical (SOC) mechanisms of the underlying sandpile driving rules.

\begin{figure}[h!]
    \centering
    \includegraphics[width=\columnwidth]{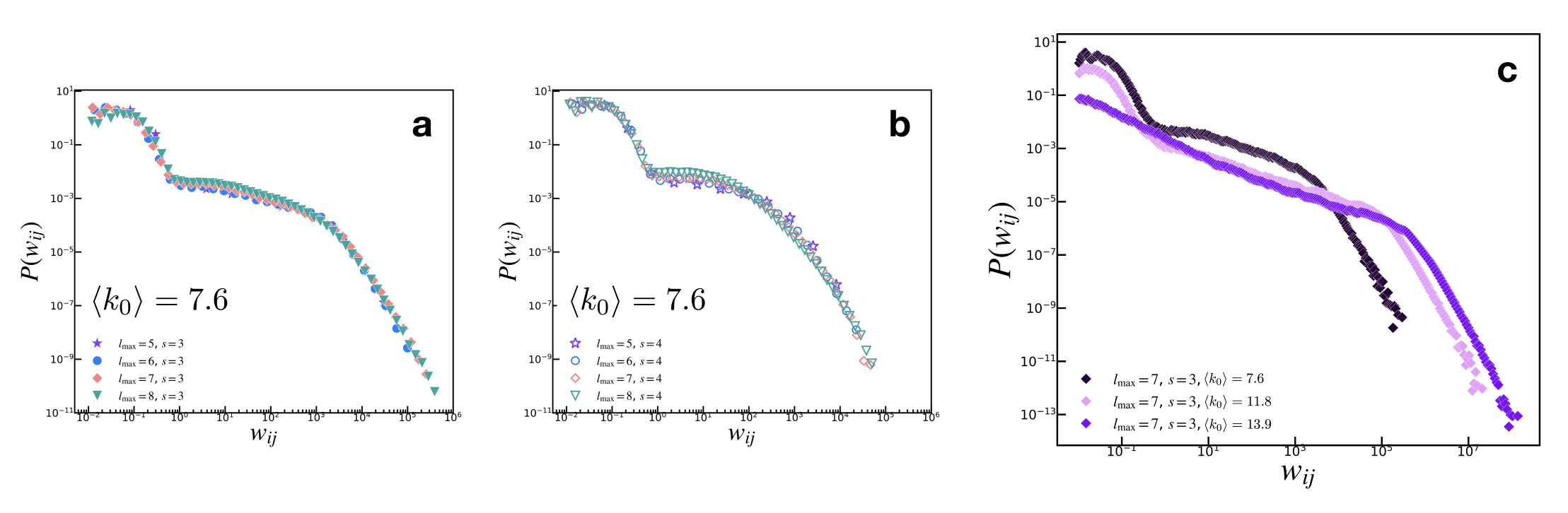}
    \caption{\label{fig:suppl-ew-e06n} \textit{Edge weight distributions for $\langle k_0 \rangle = 7.6$.} For both (a) $s=3$ and (b) $s=4$, the $P(w_{ij})$ for $\langle k_0 \rangle = 7.6$ do not produce distinct PL regimes, in contrast with (c) those generated by $\langle k_0 \rangle = 11.8$ [see also Fig.~\ref{fig:edgeweight}] and $\langle k_0 \rangle = 13.9$.
    }
\end{figure}

However, the lower density HMNs with $\langle k_0 \rangle = 7.6$ do not recover the distinct PL regimes found for the denser starting HMNs [as shown in the main text in Fig.~\ref{fig:edgeweight}, which is obtained for $\langle k_0 \rangle = 11.8$]. Even upon reaching the maximal connectivity [see \blue{ Supplementary Information: Network Evolution}], the grown networks that started with original HMNs with $\langle k_0 \rangle = 7.6$ produce non-PL distributions, as shown in Fig.~\ref{fig:suppl-ew-e06n}(a) and (b). To further highlight this limitation, Fig.~\ref{fig:suppl-ew-e06n}(c) shows the representative results for $\langle k_0 \rangle = \{7.6, 11.8, 13.9\}$ for the representative case of $l_{\max} = 7$ and $s=3$. The $\langle k_0 \rangle = 7.6$ span shorter orders of magnitude and show no clear linear trends, while the denser $\langle k_0 \rangle = 11.8$ and $\langle k_0 \rangle = 13.9$ all clearly show the two PL regimes also shown in Fig.~\ref{fig:edgeweight}(b) and (c). As such, unless otherwise specified, all of the presented results in the main text are obtained for starting HMNs with original edge density $\langle k_0 \rangle = 11.8$ or denser.

\subsection*{Scaling Behavior of Avalanche Metrics}

\begin{figure}[h!]
    \centering
    \includegraphics[width=\columnwidth]{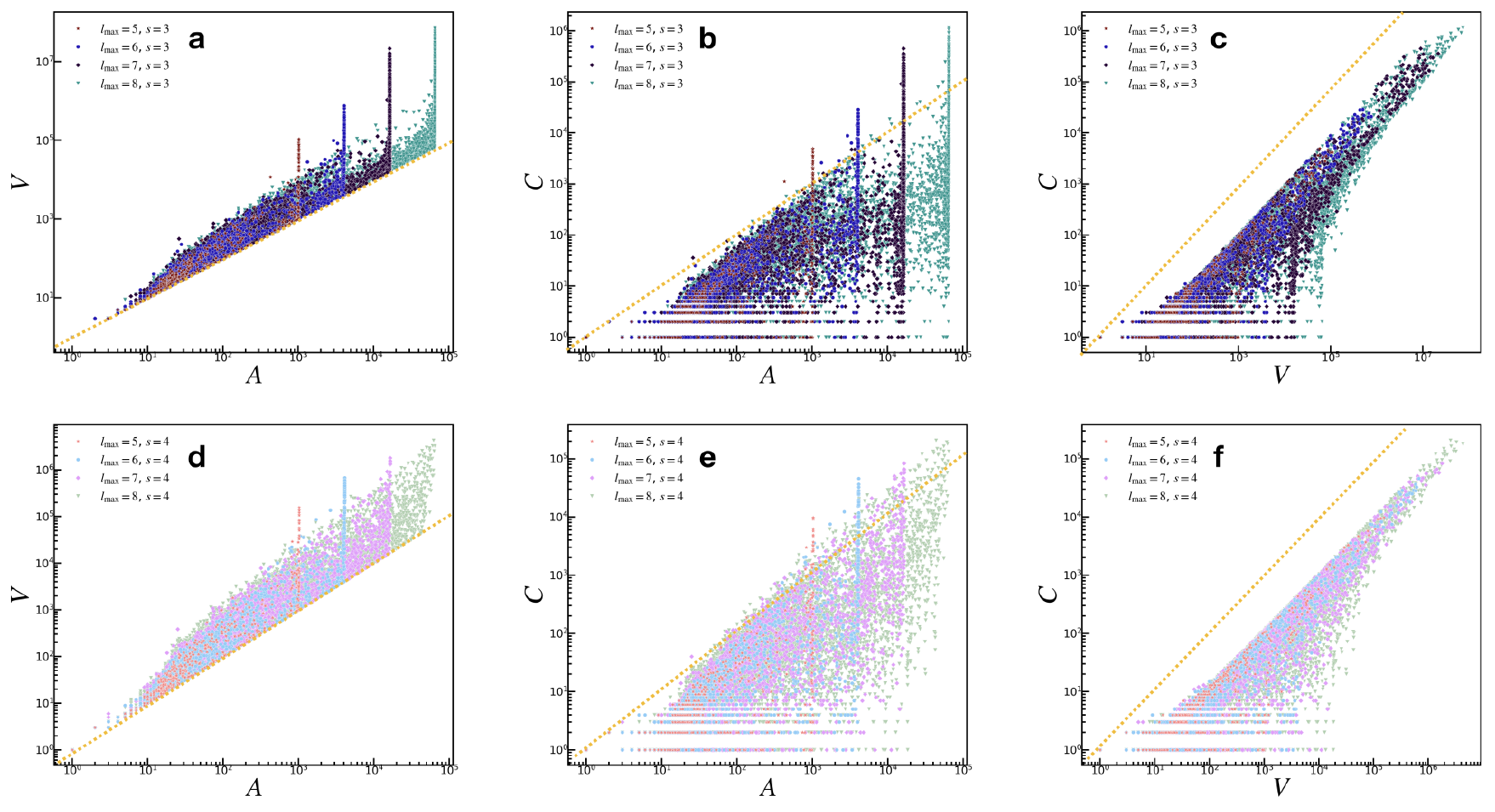}
    \caption{\label{fig:suppl-avscaling} \textit{Scaling behavior of avalanche measures} [Top panels: $s=3$; bottom panels: $s=4$; dashed line is the line of equality]. The left panels (a) and (d) scattergrams of $V$ vs. $A$ show a floor behavior corresponding to the $V=A$ line, but most points fall above this line due to multiple spatiotemporal counts done in reporting $V$. Additionally, the spikes at $A=N$ are shown for the various $l_{\max}$ considered. The middle panels (b) and (e) scattergrams of $C$ vs. $A$ fall below the $C=A$ line because the affected but non-toppled sites are not reported in $C$ but counted in $A$. Additionally, the spikes at $A=N$ are still observed. For the right panels (c) and (f) corresponding to the $C$ vs. $V$ plot, the rightward shift from the $C=V$ line is expected because only the toppled sites are counted by $C$, while all affected sites are included in $V$.}
\end{figure}

In Fig.~\ref{fig:suppl-avscaling}, every avalanche event (for each of the model parameter sets investigated) is represented by a point corresponding to its $A$, $V$, and $C$ values. All of these points are plotted in $V$ vs. $A$ [left panels], $C$ vs. $A$ [middle panels], and $C$ vs. $V$ [right panels], for both the $s=3$ [top panels] and $s=4$ [bottom panels] cases. In all plots, the straight line with slope equal to unity (corresponding to the case of equal values in the vertical and horizontal axes) are shown as the orange dashed lines.


The $V$ vs. $A$ plots in Fig.~\ref{fig:suppl-avscaling}(a) and (d) show a lower limit (floor) bounded by the $V = A$ plot, as expected; the number of spatiotemporal activations should not be lower than the number of affected nodes. The spreading of points above the $V = A$ line indicates that the spatiotemporal activity within the network can trigger individual nodes multiple times, such that $V$ can be orders of magnitude greater than $A$. In Fig.~\ref{fig:suppl-avscaling}(a) and (d), the ``area'' measures are limited by $A = N$, which is shown by the spikes at these values for the different $l_{\max}$ considered. This limitation, which is also apparent in the tails of $P(A)$ in Fig.~\ref{fig:avalanchesize}(a), indicates that the spatiotemporal activity is a better indicator of the actual effect of the avalanche on the entire network. This, in turn, is the motivation for the use of $V$ in the reinforcement mechanisms of the Hebbian learning rule, Eqn.~\ref{eqn:enhance}. 

In contrast, the scattergrams $C$ vs. $A$ in Fig.~\ref{fig:suppl-avscaling} mostly fall below the $C=A$ line especially for the smallest values of $C$ and $A$. This is because the $C$ metric takes only the sites that toppled, and not the affected ones such as in $A$. However, at the most active state of the network, the ``areas'' are still limited up to the $A=N$, while the $C$ values reach several orders of magnitude greater than $A$, indicating multiple repeated topplings. It is also interesting to note that the base of the scattergrams $C=1$ stretch throughout almost the entire $A$ axis, indicating the instances when a single toppling affected all the other nodes in the network without driving them beyond the threshold state.

Finally, the Fig.~\ref{fig:suppl-avscaling}(c) and (f) show the $C$ vs. $V$ plot, which generally follow a unit slope in the double logarithmic scale but not exactly at the line of equality $C=V$. This trend is expected because both $C$ and $V$ are measures of spatiotemporal activity. The scattergrams are shifted to the right of the $C=V$ plot due to the fact that the $V$ measures all the affected sites, while $C$ only counts those that achieved states above the threshold. Despite the strong correlation between $C$ and $V$, especially at the largest avalanches, we chose to report them separately because they report two distinct aspects of the avalanching behavior of the system. The $C$ metric is patterned after the experimental work on neural avalanches\cite{Beggs2022} that capture only the neurons that spiked; as such, the $P(C)$ distributions in Fig.~\ref{fig:avalanchesize}(c) recovered the critical exponents of $3/2$ as reported in literature. On the other hand, the $V$ is deemed to be a more detailed depiction of the actual level of activity in the network, as it counts the participation of sites that received ``grains'' even without toppling themselves. As such, the $V$ measure of avalanche activity is used for the reinforcement rule Eqn.~\ref{eqn:enhance}.

\subsection*{Robustness of Edge Weight Tail Distributions} 

The distribution of edge weights given by Fig.~\ref{fig:edgeweight} features two prominent PL regimes. The middle region of the $P(w_{ij})$ features a gentle decay exponent of around 0.6, while the tails produce steeper PLs with exponents close to 3, which is found to be universal across the different sizes and long-range connectivity of the baseline HMNs used. The steep PLs at the tails show remarkable similarities with the empirical connectome weight distributions\cite{CirunayEtAlPRR2025}, as shown in Fig.~\ref{fig:degree-datamodel}.

\begin{figure}[h!]
    \centering
    \includegraphics[width=0.8\columnwidth]{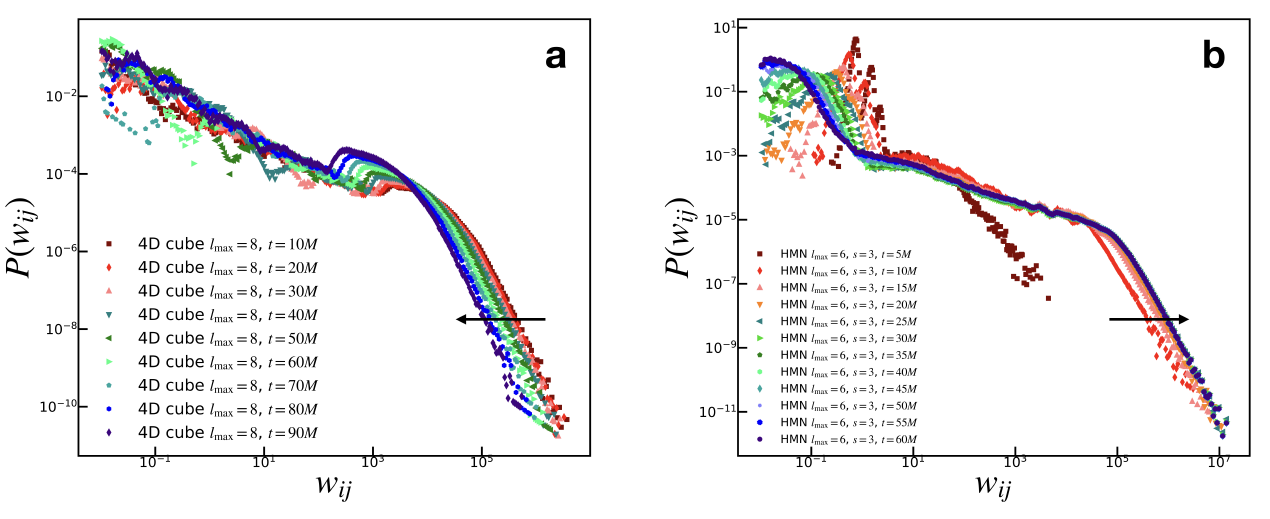}
    \caption{\label{fig:suppl-4DcubeVsHMN} \textit{Edge weight distributions for different baseline networks.} Resulting edge weight distributions for simulations that utilize (a) the four-dimensional regular lattice and (b) the HMN with the same number of nodes. Upon reaching the state of maximal connectivity, the runs on HMN show a steady approach to the universal exponent 3 tails. HMN parameters are shown by the legends. Arrows display the behavior of the distributions over longer iteration times.
    }
\end{figure}

As such, to check for the robustness of the PL regimes, especially at the tails, we compared the runs that utilized the HMNs with those that utilized simple lattice geometries. In particular, Fig.~\ref{fig:suppl-4DcubeVsHMN} shows the resulting $P(w_{ij})$ distributions for the implementations that utilized $N=4^6$ nodes initially arranged into (a) the four-dimensional lattice and (b) the HMN with $s=3$ and $\langle k_0 \rangle = 11.8$. The distributions have the same number of bins for ease in comparisons. As is apparent in Fig.~\ref{fig:suppl-4DcubeVsHMN}(b), HMNs produce $P(w_{ij})$ middle regions and the well-defined secondary mode exponent 3 PLs. On the other hand the 4d lattices (a) produce a mild middle PL region crossing over to a faster than PL decay region later. As learning commences, the distributions tend to shift leftward (a) due to the dominance of the mode that tends to shorten the tails,
or rightward (b) as dynamical criticality causes PL distributions.

The original HMN network in Fig.~\ref{fig:suppl-4DcubeVsHMN}(b) shows a strong collapse into the gentle middle region and the sharp PL tail. The collapse into the universal exponent of 3 at the tails is even more pronounced as time progresses. As noted in Fig.~\ref{fig:edgeweight}(e), after a critical iteration time, all the distributions begin to collapse, resulting in robust and nearly universal PLs. Additionally, the approach towards PLs with exponent 3 at the tails are observed for all HMN parameters sets investigated: for different network sizes $N=4^{l_{\max}}$, with $l_{\max} = \{5,6,7,8\}$; long-range connectivity, $s=\{3, 4\}$; and initial edge density, $\langle k_0 \rangle = \{7.6, 11.8\}$.


\bibliography{sample}

\section*{Acknowledgments}
We are thankful for the financial support from the Hungarian National Research, Development and Innovation Office NKFIH (Grant No. K146736 and the KIFU access to the Hungarian supercomputer Komondor.).

\section*{Additional information} 

\subsection*{Author contributions statement}

G. \'O conceptualized the work, did the theoretical analysis, and created the HMN; R.C.B. designed the model and ran the simulations; M.T.C. did the fitting with empirical data and tested the PL fits. All authors analyzed the results and reviewed the manuscript.

\subsection*{Competing interests} 

The authors declare no competing interests.

\subsection*{Data Availability} 

Data are available upon request from the authors.

\end{document}